# Tuning the electronic properties at the surface of $BaBiO_3$ thin films

C. Ferreyra[1,2,*], F. Guller[1,3,*], F. Marchini[3,4], U. Lüders[5], C. Albornoz[1], A.G. Leyva[1,6], F.J. Williams[3,4], A.M. Llois[1,3], V. Vildosola[1,3], D. Rubi[1,3,6]

1. GIyA y INN, CNEA, Av.Gral Paz 1499, (1650), San Martín, Buenos Aires, Argentina

2. Departamento de Física, Facultad Ciencias Exactas y Naturales, Universidad de Buenos Aires, Argentina.

3. Consejo Nacional de Investigaciones Científicas y Técnicas (CONICET), Argentina.

4. Departamento de Química Inorgánica, Analítica y Química-Física, INQUIMAE-CONICET, Facultad Ciencias Exactas y Naturales, Universidad de Buenos Aires, Pabellón 2, Ciudad Universitaria, Buenos Aires, Argentina.

5.CRISMAT, CNRS UMR 6508, ENSICAEN, 6 Boulevard Maréchal Juin, 14050 Caen Cedex 4, France

6. Escuela de Ciencia y Tecnología, UNSAM, Campus Miguelete, (1650), San Martín, Buenos Aires, Argentina

* These authors contributed equally to the work

Corresponding author: D.R. (rubi@tandar.cnea.gov.ar)

The presence of 2D electron gases at surfaces or interfaces in oxide thin films remains a hot topic in condensed matter physics. In particular, $BaBiO_3$ appears as a very interesting system as it was theoretically proposed that its (001) surface should become metallic if a Bi-termination is achieved (Vildosola et al., PRL 110, 206805 (2013)). Here we report on the preparation by pulsed laser deposition and characterization of $BaBiO_3$ thin films on silicon. We show that the texture of the films can be tuned by controlling the growth conditions, being possible to stabilize strongly (100)-textured films. We find significant differences on the spectroscopic and transport properties between (100)-textured and non-textured films. We rationalize these experimental results by performing first principles calculations, which indicate the existence of electron doping at the (100) surface. This stabilizes Bi ions in a 3+ state, shortens Bi-O bonds and reduces the electronic band gap, increasing the surface conductivity. Our results emphasize the importance of surface effects on the electronic properties of perovskites, and provide strategies to design novel oxide heterostructures with potential interface-related 2D electron gases.

The perovskite BaBiO$_3$ has generated a great deal of attention since the discovery of superconductivity upon doping with lead or potassium [1,2,3]. BaBiO$_3$ is a semiconducting oxide in contrast to the metallic behaviour expected for Bi$^{4+}$ ions half-filling the 6s band (this configuration implies only one Bi site per unit cell). Based on neutron diffraction experiments, the presence of two different Bi sites was established [4], related to the charge disproportionation of Bi ions (2Bi$^{4+}$→Bi$^{3+}$+Bi$^{5+}$). This charge disproportionation, together with the breathing distortions of the BiO$_6$ octahedra [4], originates the splitting of the conduction band. Therefore, the metallic behaviour of the cubic structure at high temperatures leads to a semiconducting Peierls-like phase and a monoclinic distorted structure at room temperature. Besides the fascinating properties of this material in bulk, recently, Vildosola et al., have theoretically proposed a new mechanism for the formation of a two-dimensional electron gas (2DEG) at the surface of (001) Bi-terminated BaBiO$_3$ [5]. Based on first-principle calculations, they propose that this mechanism is related to the breaking of charge ordering at the surface due to the incomplete oxygen environment of the surface ions, which renders a cubic-like metallic behaviour confined to a few monolayers close to the surface. The experimental confirmation of this effect remains very challenging and has not been achieved so far. Only a few reports dealing with BaBiO$_3$ thin films can be found on the literature. Makita et al. reported on BaBiO$_3$ thin films deposited by molecular beam epitaxy on SrTiO$_3$, showing that the orientation of their films can be controlled by growing a BaO buffer layer between the film and the substrate [6]. Later, Gozar et al. fabricated BaBiO$_3$ films by the same technique and identified their termination layer as BiO$_2$ [7]. Finally, Inumaru el at. reported on BaBiO$_3$ thin films deposited by pulsed laser deposition on MgO, and claim that, in spite that the tilting of BiO$_6$ octahedra is suppressed, their films remained "insulating" [8].

In the present work, the surface electronic and transport properties of (100)-oriented BaBiO$_3$ thin films are studied. We have grown BaBiO$_3$ thin films on Si (001) substrates and show that, although not epitaxial, the films can be highly textured in the (100) direction given the right growth conditions. X-ray photoemission experiments show the presence of an anomalous Bi contribution, which is maximized in (100) textured films. Based on new theoretical calculations, we identify that this contribution is related to electron doping at the (100) surface, which stabilizes the Bi valence in +3 states, shortens Bi-O bonds and reduces the band gap of the material when compared to the bulk case. This implies an enhanced conductivity for the (100)-surface, which is consistent with our transport measurements. These results demonstrate the importance of surface effects on the electronic properties of perovskites and suggest that combining BaBiO$_3$ with other oxides in heterostructures could be a feasible a route to produce 2DEG systems.

BaBiO$_3$ thin films were grown on top of (001) silicon by pulsed laser deposition. A Nd:YAG solid state laser, operating at λ=266nm with a repetition frequency of 10Hz, was used. The deposition temperature and oxygen pressure were 500ºC and 0.01 mbar, respectively. The used fluencies ranged between 1.9 and 3.4 J/cm$^2$. X-ray diffraction experiments were performed by means of an Empyrean (Panalytical) diffractometer with a Pixcel 3D detector. Profiles matching of the XRD spectra were done by using the FullProf software and cell parameters and volume were extracted. The film thickness was estimated by focused ion beam cross-sectioning and scanning electron microscopy imaging as 160nm. XPS measurements were performed under UHV conditions (base pressure < 5x10$^{-10}$mbar) using a SPECS UHV spectro-photometer system equipped with a 150 mm mean radius hemispherical electron energy analyser and a nine channeltron detector. Spectra were acquired at a constant pass energy of 20 eV using an un-monochromated MgKα (1253.6 eV) source operated at 12.5 kV and 20 mA and a detection angle of 30º with respect to the sample normal. Spectral backgrounds were modelled by using Shirley functions, while XPS peaks were fitted with a mixture of Gaussian and Lorentzian (70%/30%) functions. Transport measurements were performed either with Quantum Design Physical Properties Measurement or Versalab Systems, in three and four-probe configurations.

Figure 1 displays X-ray diffraction patterns corresponding to two BaBiO$_3$ films grown at fluencies of 1.9 and 3.4 J/cm$^2$, respectively. We recall that both fluences are above the minimum threshold necessary for congruent growth (<1.8J/cm$^2$ for Bi-based perovskites [9]), which indicates that the target stochiometry is fully transferred to the films. Besides a small amount of segregated BaO, produced during the first growth stages and therefore localized close to the substrate/film interface in both cases (to be reported elsewhere), no other secondary phases were found. A strong change in the BaBiO$_3$-texture is found for different fluences. The film grown at lower fluency is strongly textured in the (100) direction, while the film grown at the higher fluency is more polycrystalline-like. In the case of the textured film, we checked the absence of epitaxy by performing Φ-scans around Si (202) and BaBiO$_3$ (220) reflections, to be reported elsewhere. The presence of the native amorphous SiO$_x$ layer at the Si surface prevents the epitaxial growth, in spite of the possible structural matching between Si and BaBiO$_3$ [10]. The analysis of the XRD patterns of Figure 1 indicates that our oriented film presents an out of plane cell parameter of a=(6.18±0.01)Å, in excellent agreement with the bulk crystalline structure (a=6.181Å). A similar procedure can be done for the non-oriented film; for example, from the (00l) reflections we get c=(8.66±0.01)Å which is also consistent with the bulk value (c=8.669Å). It is also worth comparing the extracted cell volumes of bulk BBO and the non-oriented film, both values being consistent within the error of the technique ((318±2)Å$^3$ and (321±2)Å$^3$, respectively). We recall that in the case of the oriented films the lack of in-plane ordering (as determined by XRD phi-scans) does not allow obtaining in-plane cell parameters by means of 4-circle diffraction. In consequence, a direct comparison between oriented and non-oriented cell volumes is not possible. The similarity between our films cell parameters and volume (the latter only possible in the case of the non-oriented film) and bulk values suggests that both families of films (oriented and non-oriented) are strain-free and there are no significant off-stochiometries, which, in the case of existing, should necessarily impact on the films structure. The inset of Figure 1(b) shows a typical atomic force microscopy image of one of our films, where a dense arrangement of well connected micrometric grains is seen.

Figures 2 (a)-(c) show the Ba-3d and Bi-4f X-ray photoemission spectra corresponding to textured and non-textured films. In the case of the Ba-3d spectra, it is found in both cases the presence of a doublet at 781.9eV and 797.5eV, corresponding to the presence of a single species, independently of the texture of the films. On the other hand, Bi-4f spectra show the presence of two doublets: a dominant doublet at 160.2eV and 165.8eV, and a second, less intense, one at 158.1eV and 163.3eV. It is worth pointing out that photoemission experiments in bulk samples [11,12] show the presence of a single, although broad, 4f doublet, which could not be unambiguously resolved into the two components expected from the Bi charge disproportionation. In our case, it is therefore reasonable to attribute the dominant doublet to a bulk-like contribution, and the second one to the presence at the surface of Bi ions with different chemical environment. Interestingly, it is found that the integrated intensities ratio $I_{secondary}/I_{dominant}$ between both doublets changes from 0.21 for the non-textured film to 0.30 for the textured film. This clearly shows that the presence of the modified Bi chemical environment is maximized in the case of the (100) orientation.

In order to shed light on the observed experimental behaviour, we have performed ab-initio calculations of the BaBiO$_3$ (100) surface, as shown in Figure 3(a). In contrast with the (001) BaBiO$_3$ surface, which can be terminated either in Bi or Ba, in the (100) case there is no particular termination. We consider a slab composed of 17 layers that are stacked following the monoclinic crystal structure in the (100) direction. There are 9 BaBiO layers, intercalated by layers composed only by 4 O atoms. Each BaBiO layer has the same chemical composition as the one shown in Fig. 3 although they are alternatively shifted one with respect to the other by half lattice parameter in the y direction. To avoid the interaction between opposite surfaces, they are separated in the z direction by an empty space volume of 12 Å. The supercells have two inversion symmetric surfaces for simplicity. The internal atomic positions were relaxed until the forces on the atoms were below 0.02 eV/Å.

We perform first-principles Density Functional Theory (DFT) calculations using the generalized gradient approximation for the exchange and correlation potential (GGA) [13]. Standard DFT calculations using local or semilocal functionals predict a semimetallic behaviour for bulk $BaBiO_3$, however, it is well known that it presents an indirect gap whose experimental reported value goes from 0.2 eV to 1.1 eV [14]. We take care of the gap problem using the modified Becke-Johnson potential (MBJ) [15] (see supplementary information of Ref. [5]). The relaxation of the supercell is performed within the VASP code [16]. The MBJ correction is done within the WIEN2K code [17].

In Figure 3(b) we analyse the evolution of the projected Bi density of states (PDOS), layer by layer, in the simulated slab (the surface layer corresponds to the plot in the top). It can be observed that the three inner layers of the slab are semiconducting as in bulk (see, for instance, Ref [5] and its supplementary material for a description of the electronic properties in $BaBiO_3$ bulk). These layers present a charge disproportionation with one Bi ion formally behaving as $Bi^{3+}$ (black solid line) and another as $Bi^{5+}$ (red dashed line) at each layer. This charge disproportionation can be detected more clearly within the energy range (-2 eV, 2 eV) around the Fermi level set at 0eV. On the other hand, the two outermost layers are also semiconducting but they are not charge disproportionated any more. The 4 Bi ions of these two layers have a formal valence 3+. Their PDOS are plotted with a solid black line, one thick and the other one thin, for each layer, to show their resemblance. The physical origin of the breaking of the charge disproportionation is the fact that, for the simulated slab, the surface BaBiO layer does not have the four apical oxygen's, giving rise to a surface electron doping. It is important to remark that the Bi occupied bands in these two surface and subsurface layers have larger bandwidths as compared to the corresponding $Bi^{3+}$ bands of the inner layers. This effect reduces the value of the gap with respect to the bulk value. To analyse the physical reason of the gap reduction, we study the evolution of the structural distortion and charge disproportionation layer by layer of the simulated slab. In Figure 4(a) we show the evolution of the average Bi-O in-plane bond-length and in Figure 4(b) the Bader charges [18] for each type of Bi within each layer. It can be noticed that both the breathing distortions in Figure 4(a) and the charge disproportionation in Figure 4(b) are washed out for the surface and sub-surface BaBiO layers. At the inner layers the breathing distortion gives rise to a difference of around 0.15 Å in the Bi-O bondlength and the charge difference between the formal $Bi^{3+}$ and $Bi^{5+}$ is around 0.35, both values being very similar to the ones obtained for bulk $BaBiO_3$ [5]. These differences almost vanish for the two outermost layers and, in particular, the Bader charges of the Bi ions there, behave more like those of $Bi^{3+}$. The suppression of breathing distortions and charge disproportionation at the surface of the (100) film are consistent with the anomalous doublet found in the XPS experiments. Another important comment is the fact that the average Bi-O bond-lengths stabilize at a value that is smaller than those of the corresponding $Bi^{3+}$ of the inner layers. Taking into account that, both for bulk $BaBiO_3$ and for the inner layers of this slab, the Bi-$s$ and O-$p$ states conform a hybrid sp-$\sigma$ band with states in the energy range going from around -6 eV to -0.5 eV for $Bi^{3+}$, in the case of the surface and subsurface $Bi^{3+}$ ions, this bandwidth increases due to the shorter Bi-O distance. This larger bandwidth, in turn, produces a decrease in the band-gap that should induce a decrease of the surface resistance. In consequence, it can be expected for (100)-textured films a more conducting behaviour in comparison with non-textured films. This is consistent with data plotted in Figure 5(a), which shows the evolution of the resistance as a function of the temperature for both textured and non-textured films. Although both cases display a semiconducting behaviour (the resistance increases as the temperature is decreased), the textured film displays a lower resistance (around one order of magnitude at room temperature) with respect to the non-textured case. On the other hand, we recall in the case of an (100)-surface enhanced conductivity, the longitudinal (in plane) resistance of films with different thicknesses should not display significant variations. Figure 5(b) shows, for oriented films with different thicknesses, the evolution of $R_\gamma = R_{4p} \times (t/L)$, where $R_{4p}$ is the four-points resistance, t is the width of the sample and L is the distance between the voltage contacts. In the case of bulk conductivity, it would be expected a $R_\gamma$ vs. thickness dependence as that depicted with a full red line in Figure 5(b) ($R_\gamma$ scaling with 1/thickness). We experimentally find, instead, that $R_\gamma$ remains roughly constant for different thicknesses, consistently with the increased surface conductivity that our theoretical calculation suggests for (100)-oriented films. We recall that transport experiments were repeated on BBO films grown on (highly insulating) $SiO_2$ buffered Si, with similar results to what was reported above. This allows to safely discard the semiconducting substrate or the substrate/BBO interface contributing to the measured transport properties. Further experimental work involving other techniques such as conductive-atomic force microscopy are needed to get microscopic evidence on the increase of electrical conductivity at the (100) surface.

In summary, we have shown that it is possible to control the out-of-plane texture of non-epitaxial BaBiO$_3$ thin films by controlling the growth conditions. XPS spectra display an anomalous Bi doublet which is maximized for (100)-texture. The origin of this doublet is linked to our ab-initio calculations results, which suggest that electron doping stabilizes Bi$^{3+}$ species and shortens Bi-O bonds at the (100)-surface. This also implies a reduction of the band gap and an increase of the surface conductivity, consistently with our transport measurements. We conclude that as the texture of the films can be easily controlled by the growth conditions, it is possible to tune the electronic properties at the surface of these films. Finally, we would like to stress out that as this surface related effects come from the incomplete oxygen environment at surface Bi ions, a stronger effect can be anticipated if BaBiO$_3$ is combined in heterostructures with an oxide with a high oxygen affinity such as yttrium-stabilized zirconia (YSZ), which could favour oxygen transfer between both layers. In particular, the possibility of developing a metallic behaviour at such interface should be explored in the future.

We acknowledge financial support from CONICET (PIPs 291 and 273) and CIC-Buenos Aires. We thank Dr. D. Vega, from the Laboratory of X-ray Diffraction (GIA, GAIyANN, CAC, CNEA), for the XRD measurements. We thank P. Granell and Dr. F. Golmar, from INTI, for the SEM-FIB measurements, and Dr. L. Steren for useful discussions. We thank M. Moreau Linares for the AFM measurement.

Figure Captions

Figure 1: (color online) X-ray diffraction patterns corresponding to BBO thin films grown at fluences of (a) 1.9 and (b) 3.4 J/cm$^2$. Zero-shifts were corrected by using the Si reflections. A strong change in texture was found. The peak marked with * corresponds to a small amount of segregated BaO. The inset in (a) blows up both patterns for 28º<2θ<33º. The shift between (200) and (11-2) reflections is around 0.2º, well above the resolution of the diffractometer. The inset in (b) shows a typical atomic force microscopy image of one of our BBO films.

Figure 2: (color online) (a) Ba-3d X-ray photoemission spectra corresponding to both textured and non-textured BBO thin films. No significant difference between both spectra is found; (b), (c) Bi-4f X-ray photoemission spectra corresponding to textured and non-textured BBO thin films, respectively.

Figure 3: (color online) (a) (Top) BaBiO$_3$ crystal structure projected onto the xy plane. The dashed line indicates the cut made to simulate the surface; (a) (Bottom) Sketch showing the surface atoms unit cell; (b) Projected Bi ions densities of states (PDOS) at each layer. Going from the bottom to the top plots, we show the PDOS at the deepest layer successively up to the surface at the top of the Figure. The solid (black) lines correspond to the PDOS of the Bi ions behaving as Bi$^{3+}$ while the dashed (red) lines to the ones behaving as Bi$^{5+}$.

Figure 4: (color online) (a) Average Bi-O bond lengths and (b) the Bader charges, for the two types of Bi sites obtained for the simulated (100) BaBiO3 slab, at each layer. In the abscissa,

1 corresponds to the surface layer while 5 to the deepest one. Squares and circles correspond to the $Bi^{3+}$ and $Bi^{5+}$, respectively.

Figure 5: (color online) (a) Resistance as a function of the temperature for both textured and non-textured BBO films; (b) Normalized $R_\gamma$ (as defined in the main text) as a function of the thickness for (100)-oriented BBO films. The red line shows the expected evolution in the case of bulk conductivity.

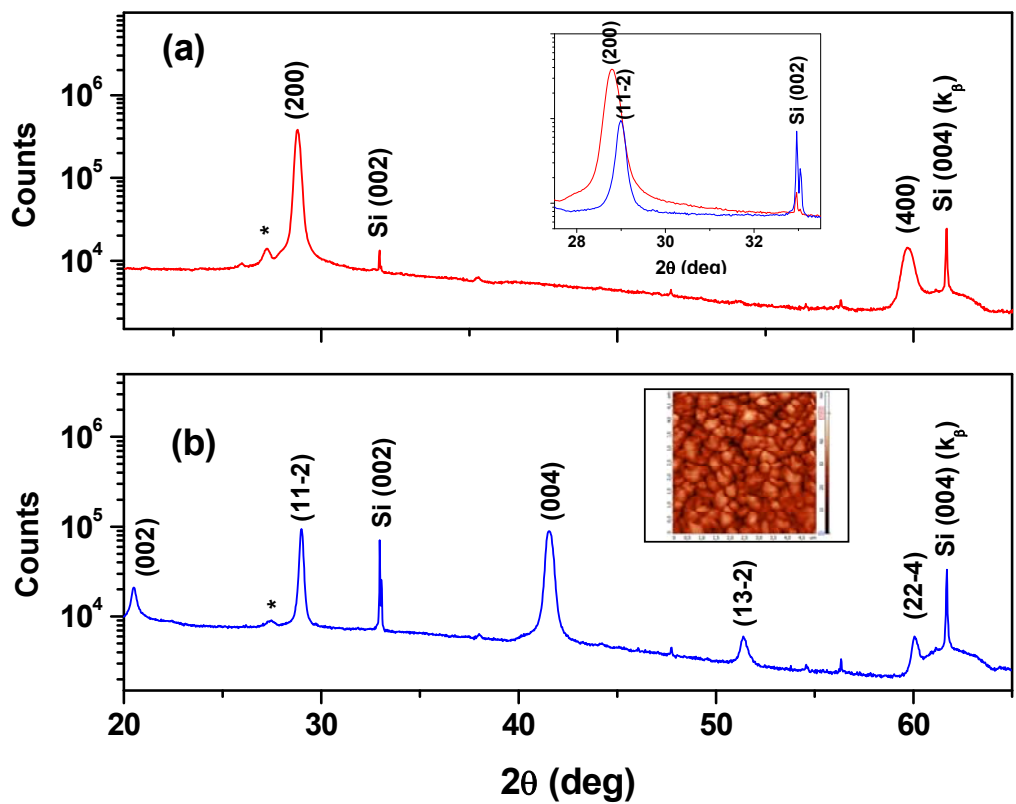

**Figure 1**

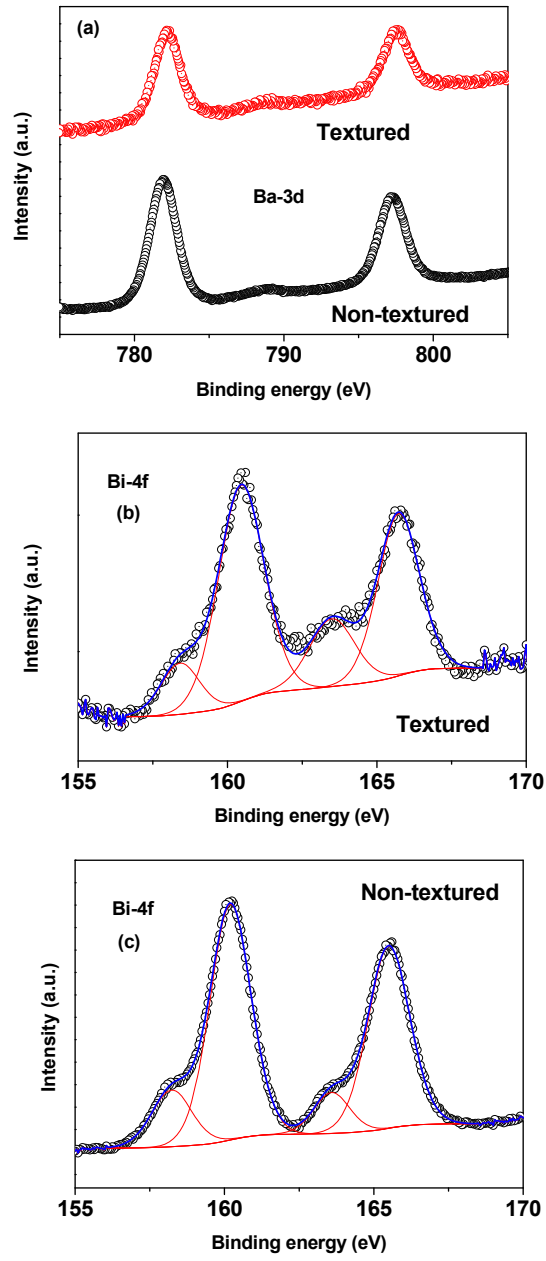

Figure 2

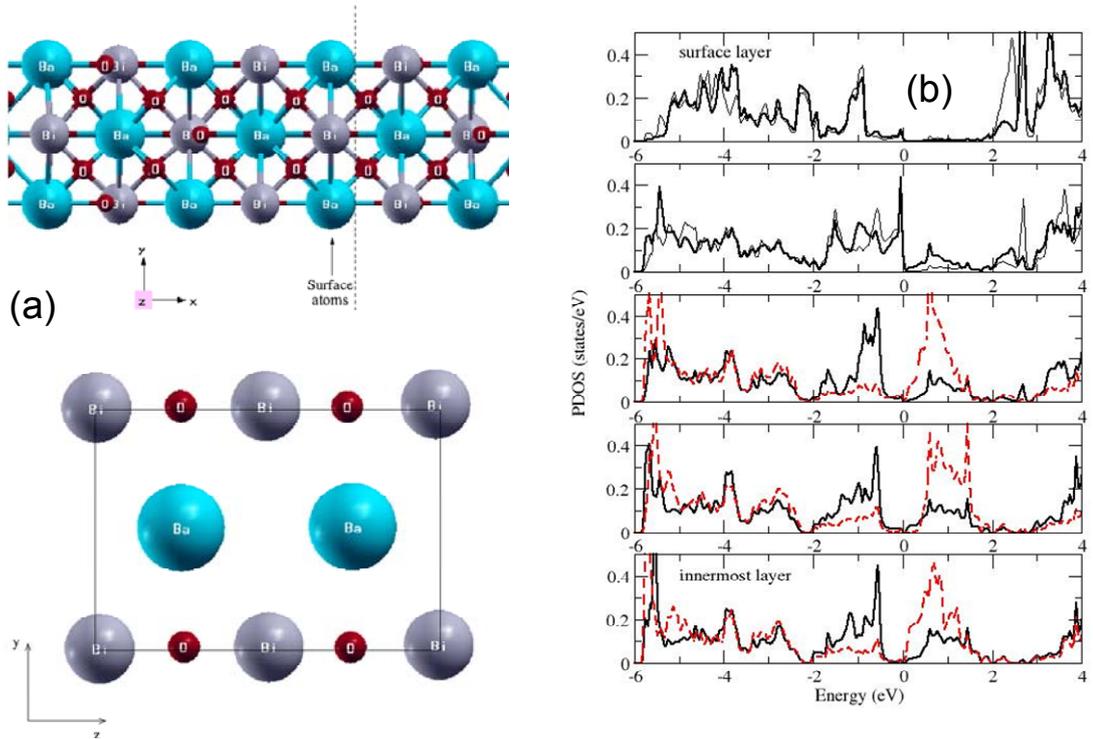

**Figure 3**

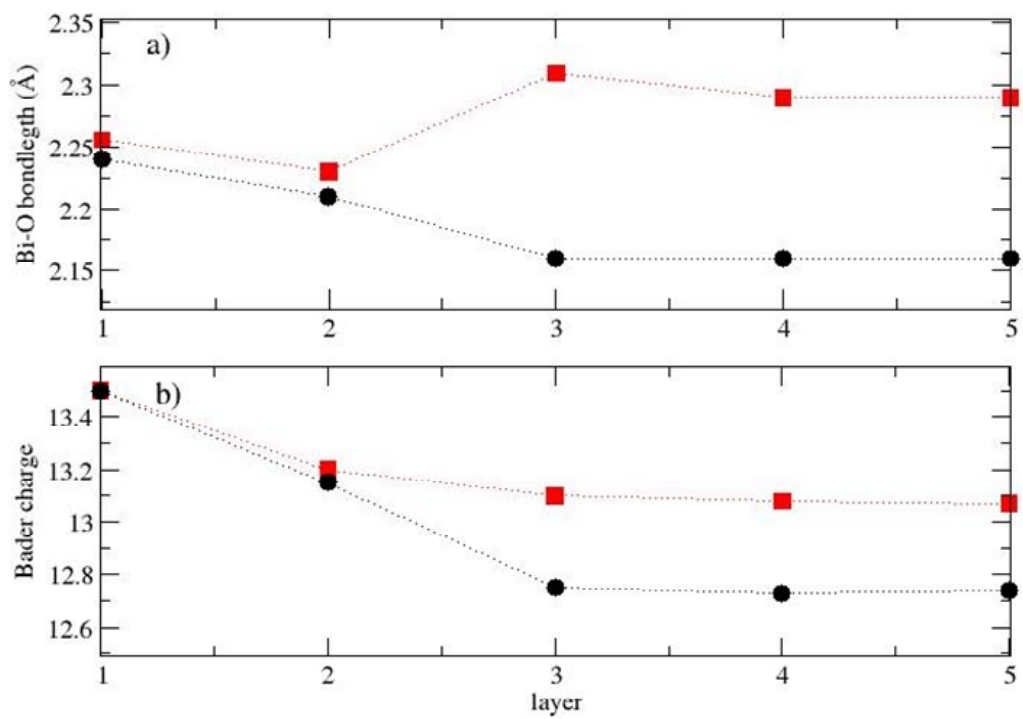

**Figure 4**

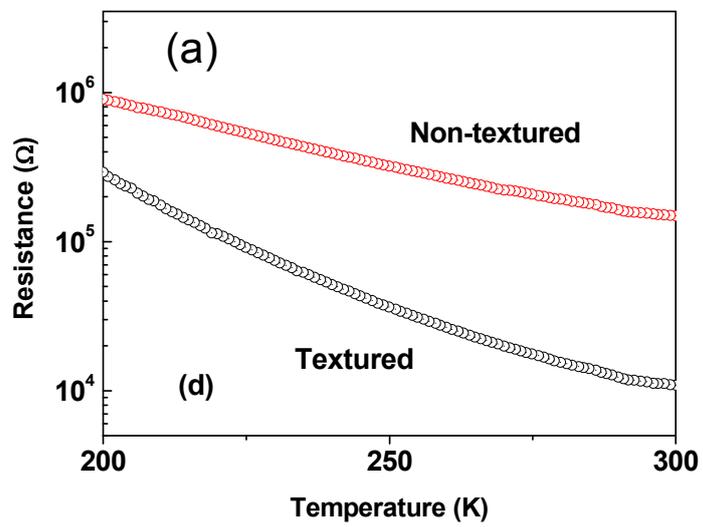

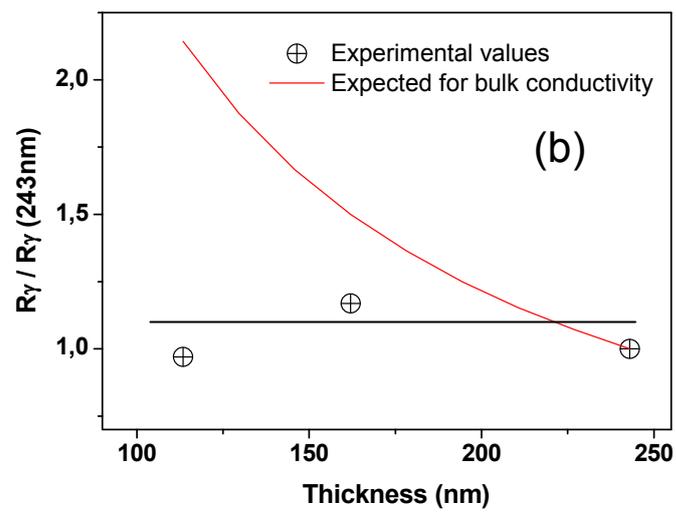

**Figure 5**